\documentclass[aps, twocolumn,prb]{revtex4}

\usepackage{amsmath}
\usepackage{graphicx}
\usepackage{epsfig}

\newcommand{\tz}{\tilde{z}}
\newcommand{\acos}{{\arccos}}
\newcommand{\nv}{{\bf n}}
\newcommand{\tv}{{\bf t}}
\newcommand{\ev}{{\bf e}}
\newcommand{\Fv}{{\bf F}}

\begin{document}

\title{The physics of stone skipping}
\author{Lyd\'eric Bocquet}
\affiliation{D\'epartement de Physique des Mat\'eriaux, UMR CNRS 5586, 
Universit\'e Lyon-I, 43 Bd du 11 Novembre 1918, 69622 Villeurbanne
Cedex, France}


\begin{abstract}
The motion of a stone skimming over a water surface is considered.
A simplified description of the collisional process
of the stone with water is proposed. The maximum number of bounces is estimated by considering 
both the slowing down of the stone and its angular stability.
The conditions for a successful throw are discussed.
\end{abstract}

\maketitle


\section{Introduction}\label{sec1}
Nearly everyone has tried to throw a stone on a lake and count
the number of bounces the stone was able to make. Of course the
more, the better.\cite{skip} Our intuition gives us some empirical
rules for the best throw: the best stones are flat and rather
circular; one has to throw them rather fast and with a small angle
with the water surface; a small kick is given with a finger to
give the stone a spin. Of course these rules can be understood
using the laws of physics: the crucial part of the motion is the
collisional process of the stone with the water surface. The water
surface exerts a reaction (lift) force on the stone, allowing it
to rebound. This process is quite complex because it involves the
description of the flow around the immersed
stone.\cite{Stong,Crane} Some energy is also dissipated during a
collision, so that after a few rebounds, the initial kinetic
energy of the stone is fully dissipated and the stone sinks.

The purpose of this paper is to propose a simplified description of
the bouncing process of a stone on water, in order to estimate the
maximum number of bounces performed by the stone. This problem
provides an entertaining exercise for undergraduate students,
with simple explanations for empirical laws that almost everyone
has experienced.

\section{Basic Assumptions}\label{hyp}
\begin{figure}[h]
\epsfig{figure=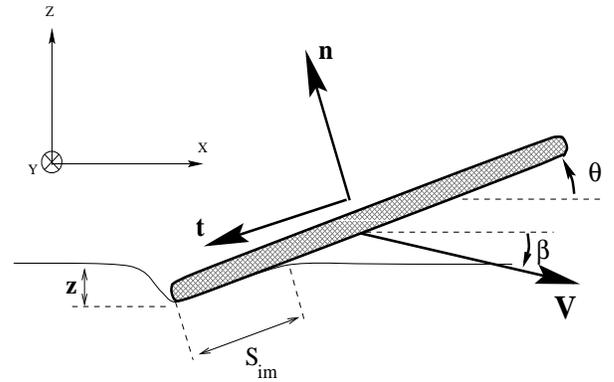,width=7.9cm,height=5cm} 
\caption{\label{fig:fig1}Schematic view of the collisional process
of a flat stone encountering a water surface. The stone has a
velocity $V$, with an incidence angle $\beta$, while $\theta$ is
the tilt angle of the stone. The immersed area $S_{\rm im}$ 
represents the area of the stone in contact with the water
surface. The depth of the immersed edge is $z$.}
\end{figure}

Consider a flat stone, with a small thickness and a mass $M$. The
stone is thrown over a flat water surface. The angle between the
stone surface and the water plane is $\theta$. A schematic view of
the collisional process is shown in Fig.~\ref{fig:fig1}. The
velocity $V$ is assumed to lie in a symmetry plane of the stone
(the plane of the paper). The difficult part of the problem is, of
course, to model the reaction force due to the water, which
results from the flow around the stone during the stone-water
contact. It is not the aim of this paper to give a detailed
description of the fluid flow around the colliding stone. Rather I
shall use a simplified description of the force keeping only the
main ingredients of the problem. First, the velocity $V$ of the
stone is expected to be (at least initially) the order of a few
meters per second. For a stone with a characteristic size $a$ of
the order of a few centimeters, the Reynolds number, defined as
${\cal R}e= Va/\nu$, with $\nu$ the kinematic viscosity ($\nu\sim
10^{-6}\,{\rm m^2\,s^{-1}}$ for water), is of order ${\cal R}e\sim
10^{5}$, that is, much larger than unity.\cite{Tritton} In this
(inertial) regime, the force due to the water on the stone is
expected on dimensional grounds to be quadratic in the velocity
and proportional to the apparent surface of the moving object and
the mass density of the fluid.\cite{Landau} Because the stone is
only partially immersed in water during the collisional process,
we expect the force to be proportional to the immersed surface
(see Fig.~\ref{fig:fig1}). The force can be adequately decomposed
into a component along the direction of the stone (that is, along
$\tv$, see Fig.~\ref{fig:fig1}) and a component perpendicular to
it (that is, along $\nv$). The latter corresponds to the lift
component of the force, and the former corresponds to a friction
component (of water along the object). I write the reaction force
due to water, $\Fv$, as:
\begin{equation}
\Fv= {1 \over 2} C_l \rho_w V^2 S_{\rm im} \nv + {1 \over 2} C_f
\rho_w V^2 S_{\rm im} \tv\, , \label{force}
\end{equation}
where $C_l$ and $C_f$ are the lift and friction coefficients,
$\rho_w$ is the mass density of water, $S_{\rm im}$ is the area of
the immersed surface, and $\nv$ is the unit vector normal to the
stone (see Fig.~\ref{fig:fig1}). 
Note that in general, both $C_l$
and $C_f$ are functions of the tilt angle $\theta$ and incidence
angle $\beta$, defined as the angle between velocity $V$ and the
horizontal. In the simplified analysis I will assume that
both $C_l$ and $C_f$ are constant and independent of tilt and
incidence angles.\cite{Note} This assumption is not a strong one
because ricochets are generally performed with a
small tilt angle, $\theta$, and a small incidence angle, $\beta$.
If one denotes the initial components of the incident velocity by
$V_{x0}$ and
$V_{z0}$ (parallel and perpendicular to the water surface,
respectively), the latter assumption amounts to $V_{z0} \ll
V_{x0}$.

We expect the lift force to be maximum when the object is
only partially immersed due to the lack of symmetry between the
two sides of the stone. Therefore, if the object reaches a depth
such that it becomes completely immersed, the lift force would be
greatly diminished and would probably not be able to sustain the
weight of the stone anymore. For simplicity, I will assume
that the lift force vanishes for completely immersed objects. The
model for the force in Eq.~(\ref{force}) is crude, but it is
expected to capture the main physical ingredients of the
stone-water interaction. It might fail for lower stone velocities
or larger incidence angles, where a bulge of water could be
created and affect the lift and friction forces on the
stone.\cite{Stong} However, in this case it is expected that the
stone will be strongly destabilized during the collision process
and perform only a very small number of bounces. We will restrict
ourselves to large initial velocities and small incidence angles,
such that the number of bounces is sufficiently large.

\section{Equations of motion}\label{equations}
Consider the collisional process, that is, the time during which
the stone is partially immersed in water. I will assume in this
section that the incidence angle $\theta$ between the stone and
the water surface is constant during the collisional process. The
validity of this assumption is considered in detail in
Sec.~\ref{spin}. The origin of time, $t=0$, corresponds to the
instant when the edge of the stone reaches the water surface.
During the collisional process, the equations of motion for the
center of mass velocity are
\begin{subequations}
\label{eom}
\begin{eqnarray}
M {dV_x\over dt} &=& - {1 \over 2} \rho_w V^2 S_{\rm im} (C_l \sin
\theta +C_f \cos \theta) \label{comx} \\ M {dV_z\over dt} &=& -Mg
+ {1 \over 2} \rho_w V^2 S_{\rm im} (C_l \cos \theta -C_f
\sin\theta)\,, \label{comz}
\end{eqnarray}
\end{subequations}
with $V^2=V_{x}^2+V_{z}^2$ and $g$ is the acceleration due to
gravity. Note that in Eq.~(\ref{eom}) the area $S_{\rm im}$
depends on the immersed depth and thus varies during the
collisional process.

Equation~(\ref{eom}) is non-linear due to the $V^2$ terms on the
right-hand side, but also due to the dependence of the immersed
area, $S_{\rm im}$, on the height $z$. However, we can propose a
simple approximation scheme: the magnitude of the velocity, $V$,
is not expected to be strongly affected by the collision process
(as I shall show in Sec.~\ref{bounces}). I thus make the
approximation that $V^2\simeq V_{x0}^2+V_{z0}^2\simeq V_{x0}^2$ on
the right-hand side of Eq.~(\ref{eom}). The validity of this
assumption requires a sufficiently high initial velocity,
$V_{x0}$, and it might fail in the last few rebounds of a stone
skip sequence.

With this approximation, Eq.~(\ref{comz}) decouples from 
Eq.~(\ref{comx}). I thus first focus the discussion on the
equation for the height $z$, which is the height of the
immersed edge (see Fig.~\ref{fig:fig1}). Note that the equation
for $z$ is equivalent to the equation of the center of mass
position, Eq.~(\ref{comz}) because $\theta$ is assumed to be
constant (see Sec.~\ref{spin} for a detailed discussion of this
point). Hence, we may identify $V_z$ with $dz/dt$ and
Eq.~(\ref{comz}) yields a closed equation for the height $z$. 

\section{Collisional process}\label{collision}
To solve Eq.~(\ref{comz}) we need to prescribe the $z$ dependence
of the immersed area $S_{\rm im}$. This quantity depends on the
precise shape of the stone. A natural choice is circular, which I
will treat in Sec.~\ref{sec:circle}. However, it is enlighting to
first consider a square shape; this shape greatly simplifies the
mathematics and already contains the basic mechanisms involved.

\subsection{A Square Stone}\label{square}
In this case, the immersed area is simply $S_{\rm im}= a \vert z
\vert/\sin \theta$ (see Fig.~\ref{fig:fig1}), with $a$ the length
of one edge of the stone. The equation for $z$ thus becomes
\begin{equation}
M {d^2z\over dt^2} = -Mg - {1 \over 2} \rho_w V_{x0}^2 C {a z
\over \sin \theta}\,, \label{Eq_z_square}
\end{equation}
where $C=C_l \cos \theta -C_f \sin\theta \simeq C_l$, and I have
used $\vert z\vert=-z$ ($z<0$). We define the characteristic
frequency $\omega_0$ as
\begin{equation}
\omega_0^2={C \rho_w V_{x0}^2 a \over 2 M \sin\theta}\,,
\label{def_omega0}
\end{equation}
and rewrite Eq.~(\ref{Eq_z_square}) as
\begin{equation}
{d^2z\over dt^2} +\omega_0^2 z=-g\,. \label{zequ}
\end{equation}
With the initial conditions at $t=0$ (first contact with water),
$z=0$ and $\dot{z} =V_{z0} < 0$, the solution of
Eq.~(\ref{zequ}) is
\begin{equation}
z(t)=-{g\over \omega_0^2} +{g\over \omega_0^2} \cos \omega_0 t +
{V_{z0}\over \omega_0} \sin \omega_0 t\,. \label{sol_z}
\end{equation}
Equation~(\ref{sol_z}) characterizes the collisional process of
the stone with water. After a collision time $t_{\rm coll}$
defined by the condition $z(t_{\rm coll})=0$ ($t_{\rm coll}\simeq
2\pi/\omega_0$), the stone emerges totally from the water surface.
It is easy to show that the maximal depth attained by the stone
during the collision is 
\begin{equation}
\vert z_{\rm max}\vert= {g \over \omega_0^2}
\left[1+\sqrt{1+\left({\omega_0 V_{z0}\over g}\right)^2}\right].
\label{zmax}
\end{equation}
As discussed in Sec.~\ref{sec1}, the stone will rebound if it
stays only partially immersed during the collision. The rebound
condition can be written as $\vert z_{\rm max}\vert < a\sin
\theta$. If we use Eqs.~(\ref{zmax}) and (\ref{def_omega0}), this
condition can be written after some straightforward calculations as
\begin{equation} V_{x0}>V_c= {\sqrt{ {4 Mg\over C\rho_w a^2}}\over
\sqrt{1-{2 \tan^2\beta M\over a^3C\rho_w\sin\theta}}}\,,
\label{condition_Vx_square}
\end{equation}
where the incidence angle $\beta$ is defined as
$V_{z0}/V_{x0}=\tan\beta$. Therefore, we obtain a {\it minimum
critical velocity} for skimming. Using the typical values,
$M=0.1$\,kg, $L=0.1$\,m, $C_l\approx C_f\approx 1$,
$\rho_w=1000$\,kg\,m$^{-3}$, and $\beta\sim\theta\sim 10^\circ$, we
obtain $V_c \simeq 0.71\,{\rm m~s^{-1}} \sim 1$\,m\,s$^{-1}$.

The physical meaning of this condition is clear: it simply
expresses the fact that the lift force ${1\over 2} C \rho_w V^2
a^2$ has to balance the weight of the stone $Mg$ in order for it to
bounce.

\subsection{A circular stone}\label{sec:circle}
For a circular stone, the immersed area is a more complex function
of the height $z$, and is given in terms of the area of a
truncated circle. A simple integral calculation yields
\begin{equation}
\label{sim} S_{\rm im}(s)=R^2 [
{\acos}(1-s/R)-(1-s/R)\sqrt{1-(1-s/R)^2}]\,,
\end{equation}
with $s=\vert z\vert/\sin \theta$ (the maximum immersed length)
and $R=a/2$ as the radius of the stone.

The equation of motion for $z$, Eq.~(\ref{comz}), thus becomes 
non-linear. However, it is possible to describe (at least
qualitatively) the collisional process and obtain the condition
for the stone to bounce.

I first introduce dimensionless variables to simplify the
calculations. The dimensionless height, $\tz$, time, $\tau$, and
immersed area, $\cal{A}$, are defined as $\tz=-z/R\sin \theta$,
$\tau= \omega_0 t$, and $ {\cal A}(\tz)=S_{\rm im}/R^2$. (The
minus sign in $\tz$ is introduced for convenience.) If we use
these variables, Eq.~(\ref{comz}), and $V_z=dz/dt$, we
obtain
\begin{equation}
{d^2\tz\over d\tau^2}=\alpha - {1\over 2} {\cal A}(\tz)\,,
\label{eom_circle}
\end{equation}
with $\alpha=g/(R\omega_0^2\sin \theta)$.
Equation~(\ref{eom_circle}) is the equation of a particle (with
unit mass) in the potential ${\cal V}(\tz)=\int ({1\over 2} {\cal
A}(\tz)-\alpha) d\tz$. We can use standard techniques for
mechanical systems to solve Eq.~(\ref{eom_circle}). In particular,
Eq.~(\ref{eom_circle}) can be integrated once to give the
``constant energy" condition
\begin{equation}
{1 \over 2} \left({d\tz\over d\tau}\right)^2+{\cal V}(\tz)=E\,,
\label{energy}
\end{equation}
where $E$ is the energy of the system and is given in terms of the
initial conditions
\begin{equation}
E={1 \over 2} \left({d\tz\over d\tau}\right)^2\vert_{\tau=0}+{\cal
V}(\tz=0)={1 \over 2} (V_{z0}/(R\omega_0 \sin \theta))^2\,.
\end{equation}

\begin{figure}[h]
\epsfig{figure=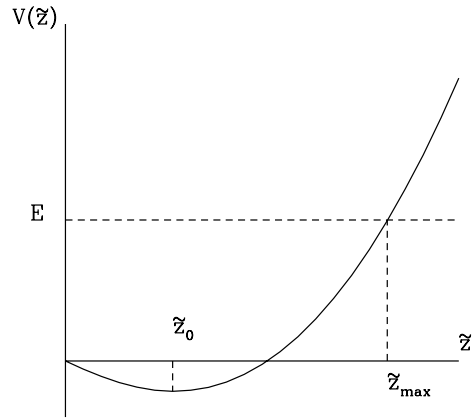,width=7.cm,height=7cm,angle=0}
\caption{\label{fig:potential}Plot of the potential ${\cal V}(\tz)$. The horizontal line is the constant energy $E$ of the system.}
\end{figure}

The potential ${\cal V}(\tz)$ can be calculated analytically using
the expression for the immersed area $S_{\rm im}$ given in
Eq.~(\ref{sim}). An integral calculation gives
\begin{eqnarray}
&{\cal V}(\tz)=&{1\over 2}\biggl(\sqrt{1-(1-\tz)^2} \left[{2\over
3}+{1\over 3} (1-\tz)^2\right] \nonumber \\
&&-(1-\tz) \acos (1-\tz)\biggr)-\alpha\tz\,.
\end{eqnarray}
This potential is plotted in Fig.~\ref{fig:potential} as a
function of $\tz$. As a consequence of the constant energy
condition, Eq.~(\ref{energy}), $\tz$ exhibits a turning point at
a maximum depth defined by ${\cal V}(\tz_{\rm max})=E$.

Here again, the condition for the stone to bounce is that this
maximum depth be reached before the stone is fully immersed, that
is, $\vert z_{\rm max}\vert < 2R \sin \theta$. In terms of
dimensionless variables, we obtain the condition: $\tz_{\rm max}
<2$, with $\tz_{\rm max}$ defined by ${\cal V}(\tz_{\rm max})={1
\over 2} (V_{z0}/(R\omega_0 \sin \theta))^2$. This condition can
be explicitly solved. Let me introduce $\tz_0$ such that $d{\cal
V}/d\tz=0$ at $\tz=\tz_0$: ${\cal V}(\tz)$ is a monotonically
increasing function of $\tz$ for $\tz>\tz_0$. Now it is easy to
show that $\tz_{\rm max}>z_0$ (because ${\cal V}(\tz_{\rm max})>0$
and ${\cal V}(\tz_0)<0$), and the condition $\tz_{\rm max} <2$ is
therefore equivalent to ${\cal V}(\tz_{\rm max}) < {\cal
V}(2)={\pi\over 2}-2\alpha$, that is, ${1 \over 2}
(V_{z0}/(R\omega_0 \sin \theta))^2<{\pi\over 2}-2g/(R
\omega_0^2\sin\theta)$. Then the condition for skimming can be
rewritten (recalling that $V_{z0}/V_{x0}=\tan \beta$)
\begin{equation}
V_{x0}> V_c={\sqrt{ {16Mg\over \pi C\rho_w a^2}}\over
\sqrt{1-{8M\tan^2\beta \over \pi a^3C\rho_w\sin\theta}}}\,.
\label{condition_Vx_circle}
\end{equation}
Up to (slightly different) numerical factors this condition is the
same as in Eq.~(\ref{condition_Vx_square}) for a square stone.
Note moreover, that the reasoning used for
the potential ${\cal V}$ is quite general and can be applied to the
square shape as well. This reasoning yields the same condition as
Eq.~(\ref{condition_Vx_square}) in this case.

Note also that for the circular stone, a simplified analysis of
the motion could have been performed. First if $\tz$ remains
small during the bounce of the stone, a small $\tz$ expansion of
${\cal V}(\tz)$ is possible, yielding ${\cal V}(\tz)= 4\sqrt{2}/15
\tz^{5/2}-\alpha \tz$ (corresponding to a parabolic approximation
for the shape of the stone near its edge). Moreover, we remark
that for small $V_{z0}$, the energy $E$ goes to zero, so that
$\tz_{\rm max}$ is defined in this case by ${\cal V}(\tz_{\rm
max})=0$. If also use the previous approximation, we obtain
$\tz_{\rm max}=(15\alpha /4 \sqrt{2})^{2/3}$. The condition for
the stone to bounce, $\tz_{\rm max} <2$, therefore yields $\alpha
< 16/15$. In terms of $V_{x0}$, this condition
gives again a minimum critical velocity for skimming, defined as
$V_c=\sqrt{\zeta M g/ C\rho_w a^2}$ with
$\zeta=15/4\simeq 3.75$. This result is thus
close to the ``exact" condition found in
Eq.~(\ref{condition_Vx_circle}) for the $V_{z0}=0$ case. 

\subsection{Energy Dissipation}\label{dissip}
I have so far described the rebound of the stone by analyzing its
vertical motion. This analysis gave a minimum velocity for
skimming which results from the balance between the weight of the
stone and the lift of the force due to water. However, some energy
is dissipated during the collision due to the ``friction"
contribution of the force (the component along $x$). This
mechanism of dissipation leads to another minimum velocity
condition, in terms of the balance between dissipation and initial
kinetic energy. Only a qualitative description of the dissipation
is given here.

As shown by Eq.~(\ref{eom}), the component $F_x$ of the reaction
force in the $x$ direction (parallel to the water surface) will
decrease the velocity of the stone. Then after a few bounces, the
condition for the stone to bounce, Eq.~(\ref{condition_Vx_square})
or Eq.~(\ref{condition_Vx_circle}), will no longer be satisfied
and the stone will stop. It is possible to estimate the decrease in
the $x$-component of the velocity using the equation for the
center of mass position, Eq.~(\ref{comx}). If we
multiply both sides of Eq.~(\ref{comx}) by $V_x$ and
integrate over a collision time, we obtain the decrease in the
kinetic energy in the $x$ direction in terms of the work of the
reaction force:
\begin{equation}
{\cal W} \equiv {1\over 2} M V_{xf}^2-{1\over 2} M V_{x0}^2 = -
\! \int_0^{t_{\rm coll}} F_x(t) V_x(t) dt\,, \label{deltaV2}
\end{equation}
where $V_{x0}$ and $V_{xf}$ are the $x$-components of the
velocity before and after the collision, $t_{\rm coll}$ is the
collision time, and $F_x={1 \over 2} \tilde{C} \rho_w V_x^2
S_{\rm im}$ is the $x$-component of the reaction force, with
$\tilde{C}=C_l \sin \theta +C_f \cos \theta$. 

A rough estimate of the right-hand side of Eq.~(\ref{deltaV2}) is 
\begin{equation}
\int_0^{t_{\rm coll}} F_x(t) V_x(t) dt \simeq V_{x0}
\int_0^{t_{\rm coll}}F_x(t) dt\,. \label{approx}
\end{equation}
Now we have the simple relation $F_x(t)=\mu F_z(t)$, with
$\mu=\tilde{C}/C$ (see Eq.~(\ref{force})). Moreover, it is
expected that the average vertical force during a collision,
$\langle F_z(t) \rangle=t_{\rm coll}^{-1} \int_0^{t_{\rm
coll}}F_z(t) dt$, is the order of the weight of the stone, $Mg$.
This point can be explicitly verified for the square stone case,
using the expression of the force $F_z$ in terms of the height
$z(t)$ and Eq.~(\ref{sol_z}). The final result is $\langle F_x(t)
\rangle \simeq \mu Mg$.\cite{Note2}. Moreover, as shown in the above (and in
particular for the square stone, although the results remain
qualitatively valid for the circular one), the collision time is
given approximatively by $t_{\rm coll}\sim2\pi/\omega_0$. We
eventually find that the loss in kinetic energy in
Eq.~(\ref{deltaV2}) is approximatively given by
\begin{equation}
{\cal W} \simeq-\mu Mg V_{x0} {2\pi \over \omega_0}=-\mu Mg
\ell\,, \label{calcul_W}
\end{equation}
where $\ell$ is defined as 
\begin{equation}
\ell=V_{x0} {2\pi \over \omega_0}=2\pi \sqrt{2M\sin\theta\over{C
\rho_w a}}\,. \label{ell}
\end{equation}
The quantity $\ell= V_x t_{\rm coll}$ is the distance along $x$
traversed by the stone during a collision. If the energy loss
${\cal W}$ is larger than the initial kinetic energy, the stone
would be stopped during the collision. Using Eq.~(\ref{deltaV2}),
this condition can be written explicitly as ${1\over 2}
MV_{x0}^2> \vert {\cal W}\vert= \mu Mg \ell$. We deduce that the
initial velocity should be larger than the minimum velocity $V_c$
in order to perform at least one bounce, that is,
\begin{equation}
V_{x0} > V_c=\sqrt{2 \mu g \ell}\,. \label{min_dissip}
\end{equation}
If we use the same numerical values as in the previous paragraph,
we obtain $\mu=1.4$, $\ell=13$\,cm, so that $V_c \approx
2$\,m\,s$^{-1}$. This criterion is more restrictive than the
previous one, Eq.~(\ref{condition_Vx_circle}). I thus consider in
the following that Eq.~(\ref{min_dissip}) is the criterion for the
stone to skim over water.

\section{Why give the stone a spin?}\label{spin}
The previous calculations assumed a constant angle $\theta$. It is
obvious that the rebound of the stone is optimized when $\theta$
is small and positive (see, for example, the value of the force
constant $C=C_l \cos\theta -C_f \sin \theta$ which decreases when
$\theta$ increases). Now, if after a collision, the stone is put
in rotation around the $y$-axis (see Fig.~\ref{fig:fig1}), that
is, $\dot\theta \neq 0$, its orientation would change by an
appreciable amount during free flight: the incidence angle
$\theta$ for the next collision has little chance to still be in a
favorable situation. The stone performs, say, at most one or two
more collisions. There is therefore a need for a stabilizing
angular motion. This is the role of the spin of the stone.

Let us denote $\dot \phi_0$ as the rotational velocity of the
stone around the symmetry axis parallel to $\nv$ in
Fig.~\ref{fig:fig1}. I neglect in the following any frictional
torque on the stone (associated with rotational motion). During
the collision, the reaction force due to the water is applied only
to the immersed part of the stone and results in a torque applied on
the stone. For simplicity, I consider only the lift part of the
force. Its contribution to the torque (calculated at the center
$O$ of the stone) can be readily calculated as ${\cal M}_{\rm
lift}=OP \cdot F_{\rm lift} \ev_y$, where $\ev_y$ is the unit vector in
the $y$ direction in Fig.~\ref{fig:fig1} and $P$, the point of
application of the lift force, is located at the center of mass of
the immersed area. This torque is in the
$y$ direction and will eventually affect the angular motion along
$\theta$. However a spin motion around $\nv$ induces a stabilizing
torque: this is the well-known gyroscopic effect.\cite{Goldstein}
The derivation of the equation of motion of the rotating object (the
Euler equations) is a classic problem and is
treated in standard mechanics textbooks (see for example,
Ref.~\onlinecite{Goldstein}). On the basis of these equations, it
is possible to derive the stabilizing gyroscopic effect. This
derivation is briefly summarized in the Appendix.

In our case, the equation for the angle $\theta$ can be written as
\begin{equation}
\ddot{\theta}+ \omega^2 (\theta-\theta_0)={{\cal M}_\theta\over
J_1}\,, \label{momentum}
\end{equation}
where $\omega=[{(J_0-J_1)/ J_1}]\dot \phi_0$, $\dot\phi_0$ is the
initial spin angular velocity (in the $\nv$ direction), and $J_0$
and $J_1$ are moments of inertia in the $\nv$ and $\tv$
directions, respectively; $\theta_0$ is the initial tilt angle and
${\cal M}_\theta=OP \cdot F_{\rm lift}$ is the projection of the torque
due to the water flow in the $y$ direction.
Equation~(\ref{momentum}) shows that in the absence of spin
motion, $\dot\phi_0=0$, the torque due to the lift force will
initiate rotational motion of the stone in the $\theta$ direction.
As discussed above, the corresponding situation is unstable. On
the other hand, spin motion induces a stabilizing torque that can
maintain $\theta$ around its initial value. The effect of the
torque can be neglected if, after a collision with the water, the
maximum amplitude of the motion of the angle $\theta$ is small:
$\delta \theta=[\theta-\theta_0]_{\rm max} \ll 1$. If we use
Eq.~(\ref{momentum}), an estimate of $\delta \theta$ can be
obtained by balancing the last two terms in Eq.~(\ref{momentum}),
yielding $\delta \theta \sim {\cal M}_\theta/(J_1 \omega^2) $
(note that up to numerical factors (${J_0-J_1)/ J_1})\sim 1$ and
$J_1\sim MR^2$, with $R$ the radius of the stone). The order of
magnitude of ${\cal M}_\theta$ can be obtained using the results
of Sec.~\ref{dissip}. The average
vertical force acting on the stone has been found to be the
order of the weight of the stone (see the discussion after
Eq.~(\ref{approx})): $\langle F_z(t)
\rangle\simeq Mg$. If we take $OP\sim R$, we obtain the simple
result ${\cal M}_\theta \sim Mg R$. The estimate for $\delta
\theta$ follows directly as $\delta \theta \sim g/(R \omega^2)$.
Therefore, the condition for $\theta$
to remain approximately constant, $\delta \theta \ll 1$, is
\begin{equation}
\dot\phi_0\sim \omega \gg \sqrt{ g\over R}\,.
\label{spin_condition}
\end{equation}
For a stone with a diameter of 10\,cm, Eq.~(\ref{spin_condition})
gives $\dot\phi_0 \gg 14$\,s$^{-1}$, corresponding to a rotational
frequency larger than a few revolutions per second ($\sim 2$\,Hz).
This condition is easily fullfilled in practice and corresponds
approximately to what we would expect intuitively for a successful
throw. Note that the condition (\ref{spin_condition}) is
independent of the center of mass velocity of the stone $V$.

\section{An estimate for the maximum number of
bounces}\label{bounces}
The estimation of the maximum number of
bounces is the most difficult and tentative part of the analysis
because many factors can in principle slow down or destabilize the
stone, some of which are extremely difficult to model (such as
irregularities of the water surface and the wind). We shall assume
the idealized situation described above (perfect surface, no wind,
idealized reaction force) and focus on two specific factors, which
appear, at least intuitively, as natural candidates for stopping
the stone.

\subsection{Slow down of the stone}
As I have discussed in Sec.~\ref{dissip}, energy is dissipated
during a collision and the $x$ component of the velocity of the
stone will decrease during each collision: after a few collisions,
all the initial kinetic energy will be dissipated. This process
can be easily formulated.

I consider a succession of $N$ collisions. Between two collisions,
the motion is parabolic (wind and air friction are neglected) and
the initial $x$ component of the velocity at the next collision is
equal to the final $x$ component of the velocity at the end of the
previous collision. The important point to note is that the energy
loss during one collision, Eq.~(\ref{calcul_W}), is independent of
the velocity $V_{x0}$ before the collision. Therefore, the
velocity of the stone after $N$ collisions obeys the relation
\begin{equation}
{1\over 2} M V_{x}^2[N]-{1\over 2} M V_{x}^2[0] = -N \mu Mg \ell
\,, \label{Ncolls}
\end{equation}
so that the stone will be stopped at a collision number $N_c$ such
that the total energy loss is larger than the initial kinetic
energy (similar to the argument leading to the critical velocity
for skimming, $V_c$, in Eq.~(\ref{min_dissip})). This criterion
corresponds to $V_{x}^2[N_c]=0$ in Eq.~(\ref{Ncolls}), and $N_c$
is given accordingly by 
\begin{equation}
N_c={V_{x}^2[0] \over 2 g {\mu} \ell}\,.
\label{number_of_collisions}
\end{equation}
If we use the same typical values as before ($M=0.1$\,kg, 
$a=0.1$\,m, $C_l \approx C_f \approx 1$, $\rho_w=1000$\,m$^{-3}$,
$\beta\sim\theta\sim 10^\circ$, we obtain $\mu\simeq 1.4$ and
$\ell\simeq 13$\,cm. We then find $N_c\approx 6$ for the initial
velocity $V_{x0}=5$\,m\,s$^{-1}$, $N_c\approx 17$ for
$V_{x0}=8$\, m\,s$^{-1}$, and $N_c \approx 38$ for
$V_{x0}=12$\,m\,s$^{-1}$. The latter number of bounces corresponds
to the world record.\cite{skip}

It is interesting to calculate the distance between two successive
collisions. As noted, the motion of the stone is
parabolic out of the water: the position \{$X,Z$\} of the particle
is given by
$X(t)=V_x t$, $Z(t)=-{1\over 2}g t^2+\vert V_z \vert t$. The next
collision will occur at a distance $\Delta X=2 V_x \vert V_z\vert
/g$. The dependence of $V_x$ on the number of collisions $N$ is
given by Eq.~(\ref{Ncolls}). On the other hand, $V_z$ does not
depend on the number of collisions because the stone rebounds
``elastically" in the $z$ direction, as follows from the analysis
of the collisional process in Sec.~\ref{collision} (see, for
example, the conservation of the energy $E$ during the collision
discussed for the circular stone). If we use Eq.~(\ref{Ncolls}),
we obtain the simple result
\begin{equation}
\Delta X[N]=\Delta X_0 \sqrt{1-{N\over N_c}}\,, \label{deltaX}
\end{equation}
where $\Delta X_0=2 V_{x0} \vert V_{z0}\vert /g$. Note that
$\Delta X_0$ is approximately equal to the distance between the
two first ricochets, $\Delta X[N=1]$, when $N_c \gg 1$. For
$V_{x0}=8$\,m\,s$^{-1}$, we obtain $\Delta X_0\approx 2.25$\, m.

\begin{figure}[h]
\epsfig{figure=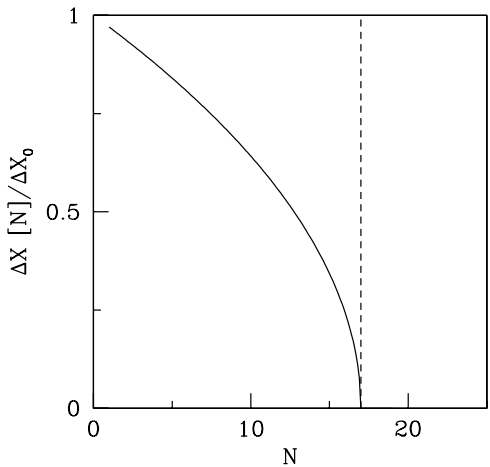,width=7.cm,height=7cm,angle=0} 
\caption{\label{fig:deltaX}Plot of the (normalized) distance between two successive collisions $\Delta X[N]/\Delta X_0$ as a function of the number of bounces $N$. The initial velocity is $V_{x0}=8$\,m\,s$^{-1}$, corresponding to $N_c=17$ (using the same values for the parameters as those given in the text). The vertical dashed line indicates that $N_c= 17$.}
\end{figure}

Equation~(\ref{deltaX}) for $\Delta X[N]$ is plotted in
Fig.~\ref{fig:deltaX}. We remark that the decrease in the distance
between two successive ricochets is first rather slow ($\Delta
X[N]\simeq \Delta X_0 (1-{N\over 2 N_c})$ for $N\ll N_c$, see
Eq.~(\ref{deltaX})), but strongly accelerates for the last
collisions when $N\sim N_c$, due to the square root variation of
$\Delta X[N]$ close to $N_c$. This result is in agreement with
observation. Such an effect is known to specialists of
stone-skipping as ``pitty-pat.''\cite{skip}

\subsection{Angular destabilization}
However, there is another possible destabilizing mechanism in the
collision process. As was discussed in Sec.~\ref{spin}, the
rotational stability of the stone is crucial in the collisional
process. A criterion for stability has been found in the form of a
minimum spin velocity of the stone. However, each collision will
perturb the rotational motion and the sum of all these effects can
eventually bypass the stability condition. This argument can be
easily formulated. As shown above, the amplitude of the angular
motion of $\theta$ is $\delta \theta \sim g/(R \omega^2)$, with
$\omega\sim\dot\phi_0$, the (constant) spin velocity of the stone.
Now assume that the destabilizing effects add, a reasonable
assumption. Then, after $N$ collisions we expect that $\Delta_N
\theta \sim N \delta\theta$. The stone is completely destabilized
for a collision number $N_c$ such that $\Delta_{N_c} \theta \sim
1$, yielding
\begin{equation}
N_c \sim {R \dot\phi_0^2\over g}\,. \label{Nc_omega}
\end{equation}
If we use the same numerical values as before, we obtain, for
example, $N_c\simeq 5$ for a initial spin velocity $\phi_0
=5$\,rev/s and $N_c=38$ (the world record\cite{skip}) for
$\phi_0=14$\,rev/s. Note, however, that there is a quite large
uncertainty of the numerical prefactors in the above estimate of
$N_c$, and this estimate is merely qualitative and should not be
taken literally. 

\section{Discussion}
At the level of our description, the maximum number of bounces
results from the combination of the two previous mechanisms: slow
down and angular destabilization. The maximum number of bounces is
therefore given by the {\it minimum} of the two previous
estimates, in Eqs.~(\ref{number_of_collisions}) and
(\ref{Nc_omega}).

The estimate $N_c^{\rm sd}$ obtained in
Eq.~(\ref{number_of_collisions}) from the slow down of the stone
depends only (quadratically) on the initial velocity of the stone:
in principle, a very large number of bounces could be reached by
increasing the initial velocity of the stone. But on the other
hand, the angular destabilization process results in a maximum
value of $N_c^{\rm spin}$ which is independent of the initial
velocity of the stone, as indicated by Eq.~(\ref{Nc_omega}). This
shows that even if the initial velocity of the stone is very
large, that is, $N_c^{\rm sd}\gg 1$, the stone will be stopped by
angular destabilization after $N_c^{\rm spin}$ bounces. In other
words, the initial ``kick" that puts the stone in rotational
motion is a key factor for a good throw.

The results presented here are in agreement with our intuition for
the conditions of a good throw. Some of the results are also in
agreement with observations, for example, the acceleration of the
number of collisions at the end of the throw (a phenomenon
known as ``pitty-pat" in stone skipping competitions \cite{skip}).
Some easy checks of the assumptions underlying
our calculations could be performed, even without any sophisticated
apparatus. For example, taking pictures of the water surface after
the ricochets would locate the positions of the collisions
(because small waves are produced at the surface of water). A
simple test of the variation of the distance between two
collisions as a function of collision number, Eq.~(\ref{deltaX}),
would then be possible. A more ambitious project 
would be to design a ``catapult,'' allowing one to throw stones
with a controlled translational and spin velocity (together with
the incidence angle of the stone on water). A measurement of the
maximum number of bounces performed for various throw parameters
would allow us to check the assumptions underlying
the present simple analysis and to determine some of the
parameters involved in the description (such as $\mu$ and $\ell$).
It would be also interesting to repeat the experiments reported in
Ref.~\onlinecite{Stong} using modern techniques (such as fast
cameras), in order to image and analyze in particular the rebound
process as a function of the throw parameters. Hopefully a better
understanding of the mechanisms of stone skipping will allow
someone to break the actual world record!

\begin{acknowledgements}
I thank my son L\'eonard for his (numerous and always renewed)
perplexing questions. I thank my colleagues from the physics
laboratory of the ENS Lyon, and in particular Bernard Castaing and
Thierry Dauxois, for their constant interest in discussing
``simple" physics problems. I am grateful to the referee for
pointing out Refs.~\onlinecite{Stong} and \onlinecite{Crane} to me.
\end{acknowledgements}

\appendix
\section{appendix}
\label{appendix}
I briefly recall the derivation of Eq.~(\ref{momentum}), from the
Euler equations described in Ref.~\onlinecite{Goldstein} The
latter are written as:\cite{Goldstein}
\begin{subequations}
\label{Euler} 
\begin{eqnarray}
& I_1 {d\omega_1 \over dt} - \omega_2 \omega_3 (I_2-I_3) &= N_1 
\label{Euler.a} \\ & I_2 {d\omega_2 \over dt} - \omega_1 \omega_3
(I_3-I_1) &= N_2 \label{Euler.b}\\ & I_3 {d\omega_3 \over dt} -
\omega_1 \omega_2 (I_1-I_2) &= N_3 \label{Euler.c}\,.
\end{eqnarray}
\end{subequations}
In Eq.~(\ref{Euler}), $I_\alpha$, $\omega_\alpha$, and $N_\alpha$
($\alpha=1,2,3$) are respectively the moment of inertia, angular
velocity, and torque along the direction of a particular principal
axis, denoted as $\alpha$. In our case, the direction 1 is taken
along the axis perpendicular to the vectors $\nv$ and $\tv$ (the
direction 1 is along the $y$ axis in Fig.~\ref{fig:fig1}), the
direction 2 along $\nv$ and the direction 3 along $\tv$. We
therefore have $\omega_1=\dot\theta$, and due to the symmetry of
the circular stone, $I_1=I_3\equiv J_1$ and $I_2\equiv J_0$.
Moreover, because only the lift component of the reaction force
(along $\nv$) is considered in the present analysis, we have
$N_1\equiv {\cal M}_\theta$ and $N_2=N_3=0$. 

Equation~(\ref{Euler.b}) yields immediately that $\dot
\omega_2=0$. We therefore have $\omega_2=\dot\phi_0$, with
$\dot\phi_0$ the initial spin velocity. Equation~(\ref{Euler.c})
can be therefore written as:
\begin{equation}
\label{thiseq} {d\omega_3 \over dt} = {J_1-J_0 \over J_1}
\dot\phi_0 \omega_1\,.
\end{equation}
If we use $\omega_1=\dot\theta$, Eq.~(\ref{thiseq}) can be
integrated once to give
\begin{equation}
\label{a3} \omega_3= {J_1-J_0 \over J_1} \dot\phi_0
(\theta-\theta_0)\,,
\end{equation}
with $\theta_0=\theta(t=0)$, the initial tilt angle. The
substitution of Eq.~(\ref{a3}) into Eq.~(\ref{Euler.a}) leads to
Eq.~(\ref{momentum}).



\newpage

\begin{thebibliography}{99}

\bibitem{skip}The actual world record appears to be 38 rebounds
(by J. Coleman-McGhee). See, for example, {\tt {<}\tt
http://www.stoneskipping.com{>}} for more information on stone
skipping competitions.

\bibitem{Stong}Some pictures of the bouncing process of a circular
stone on water and sand can be found in C. L. Stong, The Amateur
Scientist, Sci. Amer. {\bf 219} (2), 112--118 (1968).

\bibitem{Crane}H. R. Crane, ``How things work: What can a dimple
do for skipping stones?," Phys. Teach. {\bf 26} (5) 300--301
(1988).

\bibitem{Tritton} D. J. Tritton, {\sl Physical Fluid Dynamics}
(Oxford University Press, 1988), 2nd ed., pp. 97--105.

\bibitem{Landau} L. D. Landau and E. M. Lifshitz, {\sl Fluid
Mechanics} (Pergamon Press, 1959), pp. 168--175.

\bibitem{Note} Note that the nontrivial point is to assume that
$C_l$ does not vanish and reaches a finite value in the small
$\theta$ and $\beta$ limit. We may invoke the finite aspect ratio
(thickness over lateral size) of the object. For example, if the
stone is an ellipsoid of revolution with thickness $h$ and radius
$a$, with $h\ll a$, we expect $C_l\sim h/a$.\cite{Landau} However
the proportionality constant is expected to be sufficiently large
so that the lift effect is non-negligible. This property is
exemplified by water skiing. In this case, the lift force is
sufficiently large to sustain the weight of a skier on small
boards, while both tilt and incidence angles are close to zero.

\bibitem{Goldstein} H. Goldstein, {\it Classical Mechanics}
(Addison-Wesley, 1980), 2nd ed., pp. 203--213.

\bibitem{Note2} It is amusing to note that the laws of friction
for the stone are similar to those of solid friction. We have
indeed $F_x=\mu Mg$, with $\mu=\tilde{C}/C$, independent of the
velocity and surface of the stone. Of course, the same result
holds for water skiing, which is not obvious.

\end{thebibliography}
\end{document}